\documentclass[onecolumn,showpacs,amsmath,amssymb]{revtex4}
\usepackage{enumerate}
\usepackage{dcolumn}
\usepackage{graphicx}
\usepackage{bm}
\begin{document}
\title{One Local Solution of the Wheeler-DeWitt equation}

\author{Shintaro Sawayama}
 \email{shintaro@th.phys.titech.ac.jp}
\affiliation{Department of Physics, Tokyo Institute of Technology, Oh-Okayama 1-12-1, Meguro-ku, Tokyo 152-8550, Japan}
\begin{abstract}
We can obtain one solution of the Hamiltonian constraint equation in the local sense. 
The form of the state is suggested from the up-to-down method in our previous work.
The up-to-down method works for different way in treating the general metrics.
In the mini-superspace approach there appears additional constraint in the 4-dimensional quantum gravity Hilbert space.
However, in the general treatment of the metrics this method works as only solving technique.
\end{abstract}

\maketitle
\section{Introduction}\label{sec1}
There are many treatment of the quantum gravity, e.g. string theories \cite{AL} and mini-superspace approach \cite{Hart} 
or loop gravity \cite{Rov,As,Thi,As2}. The theory of the quantum gravity has not yet been constructed. 
The theory of canonical quantum gravity is based on ADM decomposition.
Form the ADM decomposition \cite{ADM}; we obtain constraint equations i.e. Hamiltonian and diffeomorphism constraint equations.
To solve these constraint equations is the orthodox way of the canonical quantum gravity.
The Hamiltonian constraint is the generator of the time translation and the diffeomorphism constraint are the generator of space translations \cite{Di2}. 
The theory of quantum gravity contains many unsolved problems which contains problem of the time and problem of the norm.
However, most important problem is the difficulty of the constraint equations, i.e. Wheeler-DeWitt equation \cite{De}.

Our motivation is simple and that is to find at least one local solution of the Wheeler-DeWitt equation.
Our work is not motivated from the higher dimensional gravity.
Although the Wheeler-DeWitt equation is difficult to solve, 
because it is the second order functional differential equation with diffeomorphism constraints, 
we had created a method to solve it that we call the up-to-down method \cite{Sa}. 
The introduced method contains many problems which is peculiar to the quantum gravity.
And some long standing problems e.g. the problems of the norm or the problem of the time has not solved yet.
In this paper we show one of the special local solutions of quantum gravity state for general spacetime metrics on the basis of the up-to-down method.
The obtained state is a solution which satisfy Hamiltonian constraint.
The main progress of the up-to-down method is the fact that we can find at least one local solution of the Hamiltonian constraint equation.

In this paper, we reconstruct a technical method to solve the Wheeler-DeWitt equation, which we call up-to-down method.
The up-to-down method consists of the following steps. 
First we add another dimension as an external time to the usual 4-dimensional metric 
and create an artificial functional space which has support of the spacetime metrics, 
and then we reduce this quantum state to the physical 4-dimensional state, 
we can simply solve the usual 3+1 Wheeler-DeWitt equation.
The same method, however does not work for Klein-Gordon systems.
The work of this method is different from the mini-superspace models, if we treat the Wheeler-DeWitt in the general sense.
If we treat the mini-superspace models, the additional constraint appears.
The simplification comes from the embedding of the 4-dimensional metric in the arbitrary 5-dimensional metric.
The ideas of the up-to-down method come from the work of dynamical horizon \cite{AK,Sa2} and the problem of the time.
The problem of the time is inversely used such that if we add an additional dimension and treat it external time, 
then this dimension or all the components of additional dimension must be vanish.
Although we use the up-to-down method as a derivation, the obtained state does not depend on the derivation. 

In section \ref{sec2} we introduces the local quantum gravity and reconstruct the up-to-down method which is solving technique of the Wheeler-DeWitt equation.
In section \ref{sec3} we derive one local solution of the Hamiltonian constraint equation without fixing spacetime metrics.
In section \ref{sec4} we summarize the obtained result and comment on the problems of the quantum gravity.
\section{Local Quantum Gravity and Up-to-down Method}\label{sec2}
The local quantum gravity introduced in this section is considered to simply treat the Wheeler-DeWitt equation.
The method of the local quantum gravity starts from decomposition of the Einstein-Hilbert action as,
\begin{eqnarray}
S=\int RdM=\sum_i \int R_i[g^{(i)}_{\mu\mu}]dS_i.
\end{eqnarray}
Here $S_i$ is the subset of the hypersurface of $\Sigma$ with constant time.
And $S_i$ is defined such that metric become diagonal by the local coordinate transformation.
The Local quantum gravity starts from decomposition of $R_i$ as usual 3+1 sense.
Then we obtained the Hamiltonian constraint and diffeomorphis constraint only in terms of the diagonal metric components.
Although the local quantum gravity only uses diagonal components of the metrics, boundary condition appears and this condition is still not well defined.

We introduce what we call the up-to-down method in self-contained way and in more strict way more than the previous paper. 
Some mistaken are corrected in this section. We should say some sentence is same as the previous paper.

We start by introducing an additional dimension which is an external euclidean time with positive signature, 
and thus create an artificial functional space corresponding to this external time.
We write such external dimension as $s$.

We dare to start with artificial 5-dimensional action whose 
metric is created from the usual 4-dimensional metric components and arbitrary additional dimensional components as,
\begin{eqnarray}
S=\int _{M\times s}{}^{(5)}RdMds.
\end{eqnarray}
Where ${}^{(5)}R$ is the 5-dimensional Ricci scalar.
Although we stat from higher dimensional action, we do not motivated from higher dimensional gravity.
Rewriting the action by a 4+1 slicing of the 5-dimensional spacetime with lapse functionals given by the $s$ direction, 
we obtain the 4+1 Hamiltonian constraint and the diffeomorphism constraints as,
\begin{eqnarray}
\hat{H}_S\equiv \hat{R}-\hat{K}^2+\hat{K}^{ab}\hat{K}_{ab} \\
\hat{H}_V^a\equiv \hat{\nabla} _b(\hat{K}^{ab}-\hat{K}\hat{g}^{ab}),
\end{eqnarray}
where a hat means 4-dimensional, 
e.g. the $\hat{K}_{ab}$ is extrinsic curvature defined by $\hat{\nabla}_a s_b$ and $\hat{K}$ is its trace, 
while $\hat{R}$ is the 4-dimensional Ricci scalar, 
and $\hat{\nabla} _a$ is the 4-dimensional covariant derivative. \\
 \\
{\it Definition. } The artificial functional state is defined by 
$\hat{H}_S|\Psi^{5} (g)\rangle =\hat{H}_V^a|\Psi^{5} (g)\rangle =0$, 
where $g$ is the 4-dimenstional spacetime metrics 
$g_{\mu\nu}$ with ($\mu =0,\cdots ,3$).
We write this functional space as ${\cal H}_5$. \\

Here, the definition of the canonical momentum $P$ is different from the usual one. 
Note in fact that the above state in ${\cal H}_5$ is not the usual 5-dimensional quantum gravity state, because the 4+1 slicing is along the $s$ direction.
It is not defined by $\partial {\cal L}/(\partial dg/dt)$ but by $\partial {\cal L}/(\partial dg/ds)$, where 
${\cal L}$ is the 5-dimensional Lagrangian.
Although whether this state is the Hilbert space or $l^2$ norm space is open question and this problem does not matter below, 
because what we would like to treat is the physical 4-dimensional quantum gravity state.

In addition, we impose that 4-dimensional quantum gravity must be recovered from the above 5-dimensional action.
The 3+1 Hamiltonian constraint and diffeomorphism constraint are,  
\begin{eqnarray}
H_S\equiv {\cal R}+K^2-K^{ab}K_{ab} \\
H_V^a\equiv D_b(K^{ab}-Kq^{ab}).
\end{eqnarray}
Here $K_{ab}$ is the usual extrinsic curvature defined by $D_at_b$ 
and $K$ is its trace, while ${\cal R}$ is the 3-dimensional Ricci scalar, 
and $D_a$ is the 3-dimensional covariant derivative.
Then we can define a subset of the auxiliary Hilbert space on which the wave functional satisfies the usual 4-dimensional constraints. 
In order to relate the 4 and 5 dimensional spaces we should define projections.
\\ \\
{\it Definition.} The subset of ${\cal H}_5$ in which the five dimensional quantum state satisfies the
extra constraints $H_SP|\Psi ^5(g)\rangle=H_V^aP|\Psi ^5(g)\rangle =0$
is called ${\cal H}_{5lim}$, where $P$ is the projection
defined by
\begin{eqnarray}
P:{\cal H}_5 \to L^2_4 \ \ \ 
\{ P|\Psi^5(g)\rangle=|\Psi^5(g_{0\mu}={\rm const})\rangle \} ,
\end{eqnarray}
where $L^2_4$ is a functional space.
And ${\cal H}_4$ is the usual four dimensional state with the restriction that 
$H_S|\Psi^4(q)\rangle=H_V^a|\Psi ^4(q)\rangle=0$.
Here $q$ stands for the 3-dimensional metric $q_{ij}(i=1,\cdots ,3)$, and
$P^{\dagger}$ is defined by
\begin{eqnarray}
P^{\dagger}:{\cal H}_{5lim} \to {\cal H}_4 .
\end{eqnarray}
The above definition contains some extra definition such as $L^2_4,{\cal H}_{5lim}$.
These functional spaces never appear in the following discussion and whether these space is norm space is open question and does not matter.
We concern that some confusion occur by these symbols. 
However, ${\cal H}_4$ should be Hilbert space and there are problems of the norm.
This problem can not have been solved yet, even if we use the up-to-down method.

From now on we consider the recovery of the 4-dimensional vacuum quantum gravity from the 5-dimensional functional.
We assume that the constraint,
\begin{eqnarray}
\hat{R}|\Psi^{5}(g)\rangle =0
\end{eqnarray}
holes. Here $\hat{R}$ is the operator,
corresponding to the usual 4-dimensional Ricci scalar.
We use this constraint for the recovery of 4-dimensional quantum gravity in the sense of Dirac.
And as we know in the previous paper physical meaning of this constraint is static restriction.
Then artificial 5-dimensional Hamiltonian constraint becomes as
\begin{eqnarray}
\hat{H}_S=-\hat{K}^2+\hat{K}^{ab}\hat{K}_{ab}:=m\hat{H}_S
\end{eqnarray}
We call this simplified 4+1 Hamiltonian constraint as modified Hamiltonian constraint.
The simplification comes from the recovery of 4-dimensional quantum gravity i.e. $\hat{R}|\Psi ^5(g)\rangle =0$.
Finally, the simplified Hamiltonian constraint in terms of the canonical representation becomes
\begin{eqnarray}
m\hat{H}_S =(-g_{ab}g_{cd}+g_{ac}g_{bd})\hat{P}^{ab}\hat{P}^{cd}.
\end{eqnarray}
The magic constant factor $-1$ for the term $g_{ab}g_{cd}$ is a consequence of the choice of the dimensions for ${\cal H}_5,{\cal H}_4$.
In the derivation of above formula we use the fact that the dimension of our universe is 4 with signature $(-,+,+,+)$.
Here $\hat{P}^{ab}$ is the canonical momentum of the 4-dimensional metric 
$g_{ab}$, that is $\hat{P}^{ab}=\hat{K}^{ab}-g^{ab}\hat{K}$. 
And as we mentioned above, this canonical momentum is defined by the external time and not by the usual time.

We now give a more detailed definition of the auxiliary 5-dimensional Hilbert space as follows: \\ \\ 
{\it Definition.} The subset ${\cal H}_{5(4)}\subset {\cal H}_5$ is defined by the constraints,  
$ \hat{R}|\Psi ^5(g)\rangle =0$, 
and we write its elements as $|\Psi ^{5(4)}(g)\rangle$.
We also define a projection $P^*$ as
\begin{eqnarray}
P^* : {\cal H}_{5(4)} \to {\cal H}_{4(5)} \ \ \  \{ P^*|\Psi ^{5(4)}(g)\rangle =|\Psi ^{5(4)}(g_{0\mu}={\rm const})\rangle 
=: |\Psi ^{4(5)}(q)\rangle \} ,
\end{eqnarray} 
where ${\cal H}_{4(5)}$ is a subset of ${\cal H}_4$.
There are reason why we choose such projection $P^*$.
Although there are the problem of measure, this definition of the projection does not produce any additional constraint in ${\cal H}_4$
for the solution of $|\Psi^{5(4)}(g)\rangle$.
So the modified Hamiltonian constraint and the 4+1 diffeomorphism constraint does not become additional constraint if they are projected by $P^*$.
\\ \\ 
{\it Definition.} 
In ${\cal H}_{4(5)}$ there is a subset whose state satisfy $H_S|\Psi^{4(5)}(q)\rangle=H_V|\Psi ^{4(5)}(q)\rangle=0$. 
We write such Hilbert 
space as ${\cal H}_{4(5)phys}$ and its elements as $|\Psi_{phys}^{4(5)}(q)\rangle$. 
If there are relations $P^*|\Psi^{5(4)}(g)\rangle=|\Psi_{phys}^{4(5)}(q)\rangle$, we write such 
$|\Psi^{5(4)}(g)\rangle$ as $|\Psi^{5(4)}_{phys}(g)\rangle$ 
and we write such Hilbert space as ${\cal H}_{5(4)phys}$.
\\ \\
We can summarize the procedure to solve the usual Wheeler-DeWitt equation.\\
1) Solve $m\hat{H}_SP^*|\Psi^4(g)\rangle =0$ and obtain $|\Psi ^{4(5)}(g)\rangle$.\\
2) Solve $H_S|\Psi ^{4(5)}(q)\rangle =H_V^a|\Psi^{4(5)}(q)\rangle =0$ to obtain $|\Psi ^{4(5)}_{phys}(q)\rangle$.\\ \\
\begin{figure}
\includegraphics{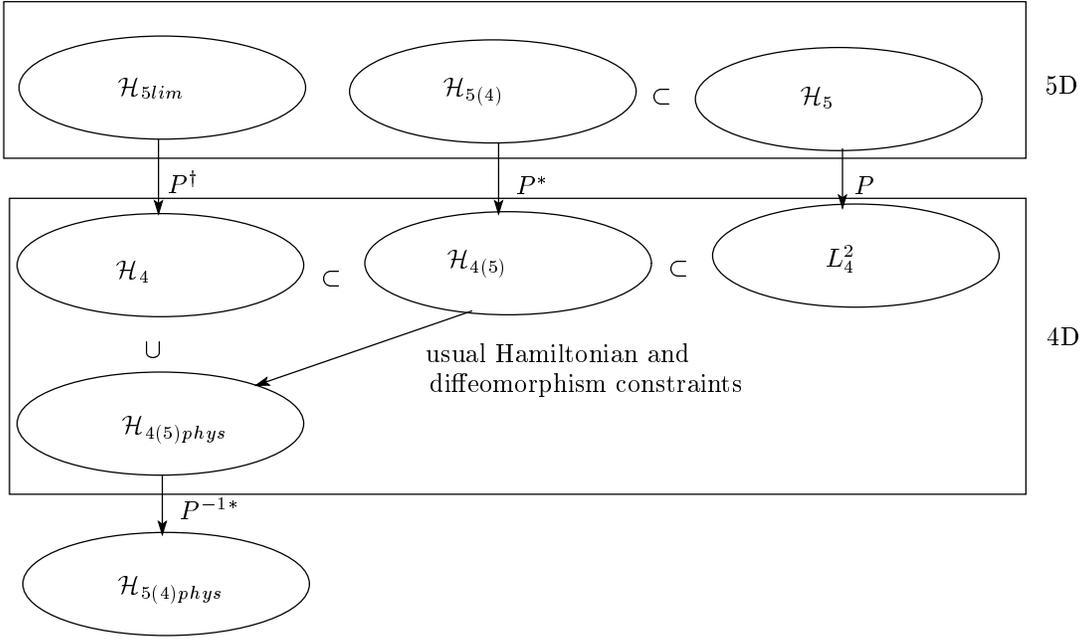}
\caption{The relations between the functional spaces and the physical 4-dimensional Hilbert space.
The Hilbert space ${\cal H}_{4(5)phys}$ is always subset of the physical Hilbert space ${\cal H}_4$.
The $P$ and $P^*$ are same operation. However, projected space is different.
The inverse projection $P^*$ can be defined by one-to-many way.
Although it is mathematically not well-defined, if we can find at least one enlargement, this method is correct.}
\end{figure}

We carry on step 1). At first we choose coordinate such that metric become diagonal because we treat local quantum gravity.
Then modified Hamiltonian constraint at this limit becomes as,
\begin{eqnarray}
m\hat{H}_SP^*=\sum_{a\not= b}q_{ii}q_{jj}\frac{\delta}{\delta q_{ii}}\frac{\delta}{\delta q_{jj}}=0.
\end{eqnarray}
We can easily find the solution of this equation and one of the states is,
\begin{eqnarray}
|\Psi^{4(5)}(q)\rangle =f_1[g_{11}]+f_2[g_{22}]+f_3[g_{33}].
\end{eqnarray}
Here $f_{i}[g_{ii}]$ for $i=0,\cdots 3$ are functional of $g_{ii}$.
Moreover, if we use a 3+1 diffeomorphism constraints, we can know that the $f_{i}$ are only the function of $g_{ii}$ at this point.
However, we do not use this fact now. 

\section{One solution of the Hamiltonian constraint}\label{sec3}
In this section we solve the local Hamiltonian constraint equation without fixing spacetime metrics, i.e. without treating mini-superspace models.
Although this section motivated by the up-to-down method, what we derive in this section independently holeds.
We use the fact that the metric becomes diagonal in the local coordinate.
We start with assumption as,
\begin{eqnarray}
|\Psi^4(q)\rangle =f_1[q_{11}]+f_2[q_{22}]+f_3[q_{33}]
\end{eqnarray}
with
\begin{eqnarray}
\hat{q}_{ii,k}f_j[q_{jj}]=\delta_{ij}q_{ii,k}f_j[q_{jj}] \ \ \ {\rm for} \ \ \ j\not= i.
\end{eqnarray}
The parameter separated solution is the superposition of three local solutions.
The assumption (16) means that the eigenvalue of the $q_{jj}$ of the state $f_i[q_{ii}]$ for $j\not=i$ is constant.
The local Hamiltonian constraint equation can be written as,
\begin{eqnarray}
\sum_{ij}\frac{1}{2}\frac{\delta^2}{\delta \phi_i\delta \phi_j}+\sum_{i\not= j}(\hat{\phi}_{i,jj}+\hat{\phi}_{j,i}\hat{\phi}_{i,i})e^{\hat{\phi}_i}=0,
\end{eqnarray}
where $q_{ii}=e^{\phi_i}$.
The consistency of the additional constraint and the Hamiltonian constraint is clear from the next commutation relation.
\begin{eqnarray}
\bigg[ \sum_{k\not= l}\frac{\delta ^2}{\delta \phi_k\delta \phi_l},
\sum_{ij}\frac{1}{2}\frac{\delta^2}{\delta \phi_i\delta \phi_j}
+\sum_{i\not= j}(\hat{\phi}_{i,jj}+\hat{\phi}_{j,i}\hat{\phi}_{i,i})e^{\hat{\phi}_i}\bigg] \sum_i f_i[\phi_i]
=\bigg[ \sum_{k\not= l}\frac{\delta ^2}{\delta \phi_k\delta \phi_l},\sum_{i\not= j}\hat{\phi}_{j,i}\hat{\phi}_{i,i}e^{\hat{\phi}_i}\bigg]\sum_i f_i[\phi_i] \nonumber \\
=\sum_{k\not= l}\frac{\delta ^2}{\delta \phi_k\delta \phi_l}\sum_{i\not= j}\hat{\phi}_{j,i}\hat{\phi}_{i,i}e^{\hat{\phi}_i}\sum_i f_i[\phi_i ] .
\end{eqnarray}
If the term $\hat{\phi}_{j,i}\hat{\phi}_{i,i}e^{\hat{\phi}_i}$ vanish, we can carry on simultaneous quantization.
And this relation is same as the equation (16), so the assumption (16) is consistent.
If we insert the equation (15), we obtain the following differential equation as,
\begin{eqnarray}
\frac{1}{2}\frac{\delta^2}{\delta \phi_i^2}f_i[\phi_i] 
+\sum_{j(\not=i)}\phi_{i,jj}e^{\phi_i}f_i[\phi_i]=0 \ \ \ {\rm for} \ \ \ i=1,2,3.
\end{eqnarray}
Other components are dropped out because of the assumption (16).
The important point is the fact that the above equation is algebraically closed for $\phi_i$.
Because we ignore the operator ordering, we can rewrite the above equation as,
\begin{eqnarray}
\frac{\delta^2}{\delta a_i^2}+4\partial^j\partial_j\ln \hat{a_i} &=&
\frac{\delta^2}{\delta a_i^2}+2i(\partial^j\partial_j\ln \hat{a_i})^{1/2}\frac{\delta}{\delta a_i} 
-2i(\partial ^j \partial _j \ln \hat{a_i})^{1/2}\frac{\delta}{\delta a_i}  
+4\partial^j \partial_j \ln \hat{a_i} \nonumber \\
&=&\frac{\delta^2}{\delta a_i^2}+2i(\partial^j\partial_j\ln \hat{a_i})^{1/2}\frac{\delta}{\delta a_i} 
-2i\frac{\delta}{\delta a_i}(\partial ^j \partial _j \ln \hat{a_i})^{1/2} 
+4\partial^j \partial_j \ln \hat{a_i} \nonumber \\
&=&\bigg( \frac{\delta}{\delta a_i}+2i(\partial^j\partial_j \ln \hat{a_i})^{1/2}\bigg)
\bigg( \frac{\delta}{\delta a_i}-2i(\partial^j\partial_j\ln {\hat a_i})^{1/2}\bigg).
\end{eqnarray}
Here, $a_i=q_{ii}^{1/2}$.
We can find the solution of the above second order ordinal functional differential equation as,
\begin{eqnarray}
f_i[g_{ii}]=E_{1i}\exp (2i\int (\partial^j\partial_j\ln a_i)^{1/2}\delta a_i)
+E_{2i}\exp (-2i\int (\partial^j\partial_j\ln a_i)^{1/2}\delta a_i) .
\end{eqnarray}
Here $E_{1i}$ and $E_{2i}$ are constants.
Then we can obtained the local Hamiltonian constraint as,
\begin{eqnarray}
|\Psi^4(q)\rangle =\sum_iE_i\cos (2\int (\partial^j\partial_j\ln a_i)^{1/2}\delta a_i).
\end{eqnarray}
We can say local solution of the Hamiltonian constraint is the cosine wave.
And above equation is the main progress of our work.

\section{Conclusion and discussions}\label{sec4}
We can find at least one local solution of the Hamiltonian constraint.
Before the up-to-down method we do not know any solution of this equation.
The peculiar problem in the quantum gravity is still unsolved.
Even if we use this method we can still not solve the problem of the norm and problem of the time \cite{Ha1,Ha2}.
We should find all state to solve the problem of norm.
Otherwise we obtained one solution it may create other solutions.
We hope such kinds of work succeed.

Because our work succeeded the quantization of the inhomogeneous spacetime, 
we can enter this state symmetry for example spherical symmetry.
Then we may obtain the state of the black holes \cite{K2}.
Or we can enter symmetry of the homogeneous and isotropy, as loop gravity \cite{Bo}.
Simple analysis is carried out by inserting symmetry in $q_{ii}$ such that $q_{11}=q_{22}=q_{33}=a$, 
where the $a$ is the function of only the time $t$.
Then we can obtain a simple cosine wave.
Such cosine wave does not occur if the cosmological constant is zero.
We can not find such wave from the mini-superspace approach.

In this method we have a problem of the matter; how we enter the matter or cosmological constant in the up-to-down method is open problem.
And this is the future work of ours.

\end{document}